\documentclass[12pt,a4paper,final]{iopart}
\usepackage{iopams}  
\usepackage{graphicx}

\begin{document}

\title[Effect of Chain Stiffness on the Structure of Single-Chain Polymer Nanoparticles]{Effect of Chain Stiffness on the Structure of Single-Chain Polymer Nanoparticles}

\author[cor1]{Angel J. Moreno$^{1,2}$}
\address{$^1$Centro de F\'{\i}sica de Materiales (CSIC, UPV/EHU) and Materials Physics Center MPC, Paseo Manuel de Lardizabal 5, 20018 San Sebasti\'{a}n, Spain.}
\address{$^2$Donostia International Physics Center, Paseo Manuel de Lardizabal 4, 20018 San Sebasti\'{a}n, Spain.}
\ead{angeljose.moreno@ehu.es}

\author{Petra Bacova$^{1,3}$}
\address{$^1$Centro de F\'{\i}sica de Materiales (CSIC, UPV/EHU) and Materials Physics Center MPC, Paseo Manuel de Lardizabal 5, 20018 San Sebasti\'{a}n, Spain.}
\address{$^3$Institute of Applied and Computational Mathematics (IACM), Foundation for Research and Technology Hellas (FORTH), 71110 Heraklion, Crete, Greece}

\author{Federica Lo Verso$^{2,4}$}
\address{$^2$Donostia International Physics Center, Paseo Manuel de Lardizabal 4, 20018 San Sebasti\'{a}n, Spain.}
\address{$^4$Department of Physics, Chemistry and Pharmacy, University of Southern Denmark, Campusvej 55, 4230 Odense, Denmark}

\author{Arantxa Arbe$^{1}$}
\address{$^1$Centro de F\'{\i}sica de Materiales (CSIC, UPV/EHU) and Materials Physics Center MPC, Paseo Manuel de Lardizabal 5, 20018 San Sebasti\'{a}n, Spain.}

\author{Juan Colmenero$^{1,2,5}$}
\address{$^1$Centro de F\'{\i}sica de Materiales (CSIC, UPV/EHU) and Materials Physics Center MPC, Paseo Manuel de Lardizabal 5, 20018 San Sebasti\'{a}n, Spain.}
\address{$^2$Donostia International Physics Center, Paseo Manuel de Lardizabal 4, 20018 San Sebasti\'{a}n, Spain.}
\address{$^5$ Departamento de F\'isica de Materiales, Universidad del Pa\'is Vasco (UPV/EHU), Apartado
1072, E-20800 San Sebasti\'an, Spain.}

\author{Jos\'{e} A. Pomposo$^{1,5,6}$}
\address{$^1$Centro de F\'{\i}sica de Materiales (CSIC, UPV/EHU) and Materials Physics Center MPC, Paseo Manuel de Lardizabal 5, 20018 San Sebasti\'{a}n, Spain.}
\address{$^5$ Departamento de F\'isica de Materiales, Universidad del Pa\'is Vasco (UPV/EHU), Apartado
1072, E-20800 San Sebasti\'an, Spain.}
\address{$^6$IKERBASQUE Basque Foundation for Science, Alameda de Urquijo 36, 48011 Bilbao, Spain.}

\newpage

\begin{abstract}
	
Polymeric single-chain nanoparticles (SCNPs) are soft nano-objects synthesized
by purely intramolecular cross-linking of single polymer chains.
By means of computer simulations, we investigate the conformational
properties of SCNPs as a function of the bending stiffness
of their linear polymer precursors.
We investigate a broad range of characteristic ratios from the fully flexible case to
those typical of bulky synthetic polymers. Increasing stiffness hinders bonding
of groups separated by short contour distances and increases looping over longer distances,
leading to more compact nanoparticles with a structure of highly interconnected loops.
This feature is reflected in a crossover in the scaling behaviour of several structural observables.
The scaling exponents change from those characteristic for Gaussian chains or rings in $\theta$-solvents
in the fully flexible limit, to values resembling fractal or `crumpled' globular behaviour for very stiff SCNPs.
We characterize domains in the SCNPs. These are weakly deformable regions that can be seen as disordered
analogues of domains in disordered proteins. Increasing stiffness leads to bigger and less deformable domains. Surprisingly, the scaling behaviour of the domains is in all cases
similar to that of Gaussian chains or rings, irrespective of the stiffness and degree of cross-linking.
It is the spatial arrangement of the domains which determines the global 
structure of the SCNP (sparse Gaussian-like object or crumpled globule).
Since intramolecular stiffness can be varied through the specific chemistry of the precursor or
by introducing bulky side groups in its backbone, our results propose a new strategy to tune
the global structure of SCNPs.

\end{abstract}

\vspace{2pc}

\section{Introduction}

A growing interest is being devoted in recent years to the synthesis of polymeric single-chain nanoparticles (SCNPs)
\cite{Altintas2016,lyon2015,reviewcsr,mavila2016,hanlon2016,PomposoSCNPbook}
with potential applications in, e.g.,  nanomedicine \cite{hamilton2009,sanchez2013design},  bioimaging \cite{perezbaena2010,bai2014}, biosensing \cite{gillisen2012single}, catalysis \cite{terashima2011,perez2013endowing,huerta2013,tooley2015}, or
rheology \cite{Mackay2003,Arbe2016,Bacova2017}.
SCNPs are obtained, generally at highly diluted conditions, through purely intramolecular cross-linking of single polymer precursors.
A series of investigations by small-angle neutron (SANS) and X-ray scattering (SAXS) \cite{sanchez2013design,perez2013endowing,Sanchez-Sanchez2013a,Moreno2013} have revealed that
SCNPs synthesized through coventional routes are usually sparse objects with open topologies.
A compilation of literature results for SCNPs in solution with very different 
chemical structures \cite{Pomposo2014a} reveals
scaling behavior, $R\sim N^{\nu}$, of the macromolecular radius $R$ with the polymerization degree $N$, 
with an average exponent $\nu \approx 0.5$. Thus, the SCNP conformations in good solvent are closer to linear
chains or rings in $\theta$-solvent
($\nu = 1/2$) \cite{Rubinstein2003} than to globular objects ($\nu = 1/3$).
As revealed by simulations \cite{Moreno2013,LoVerso2014,Rabbel2017}, the formation of long loops in the SCNP is unfrequent due to the universal self-avoiding character of the linear precursors in the good solvent conditions of synthesis.
Self-avoiding precursors actually promote bonding of reactive monomers separated by short contour distances, this
mechanism leading to the formation of small local globules (in analogy to $\theta$-solvents) but being inefficient for the global compaction of the SCNPs.
Although the cross-linking process of identical polymer precursors produces topologically polydisperse SCNPs, the resulting distribution is dominated by sparse morphologies \cite{Moreno2013,LoVerso2014,Formanek2017}.

A recent work combining simulations and SANS \cite{Moreno2016JPCL} has revealed interesting analogies
between SCNPs and intrinsically disordered
proteins (IDPs), as similar scaling behaviour  
in dilute conditions ($\nu \approx 0.5$) \cite{marsh2010,hofmann2012,bernado2012,wendell2014}
and topological polydispersity. Despite the lack of ordered regions, 
SCNPs still show weakly deformable compact `domains' (disordered analogues of the IDP domains)
connected by flexible disordered segments. The characterization of the different domains
in a given SCNP provides a structural criterion to quantify its degree of internal disorder \cite{Moreno2016JPCL}.
As a consequence of their molecular architecture with permanent loops, 
SCNPs exhibit a peculiar collapse behaviour
in concentrated solutions and melts. Instead of the Gaussian conformations
displayed by linear chains, they adopt fractal or `crumpled' globular conformations \cite{Grosberg1988,Mirny2011} resembling
those found for melts of ring polymers or chromatin loops \cite{Mirny2011,Halverson2011, Reigh2013, Goossen2014,Halverson2014}, 
suggesting this scenario for the effect
of purely steric crowding on IDPs in cell environments.

By means of simulations, in this article we extend the structural characterization of SCNPs at high dilution
by systematically investigating the effect of intramolecular bending
stiffness on the conformational properties of SCNPs and their domain structure
(previous works have been performed in the limit of flexible chains \cite{Moreno2016JPCL}). 
We find that increasing stiffness disfavours bonding
of groups separated by short contour distances. As a consequence a higher fraction of bonds at long contour distances, 
in comparison to the flexible case,  is needed to complete
cross-linking. This mechanism leads to the formation of more compact nanoparticles with a structure of highly interconnected loops. We characterize several structural observables and analyze their scaling behaviour.
The scaling exponents show a crossover from values characteristic for Gaussian chains or rings in the fully flexible limit, to values resembling crumpled globular behaviour \cite{Grosberg1988,Mirny2011} in the case of very stiff SCNPs.
We also characterize the effect of stiffness in the SCNP domains.
Increasing stiffness leads to bigger and less deformable domains though,  surprisingly, 
it has no significant effect on their scaling behaviour, which is in all cases similar
to that of Gaussian chains or rings. This suggests that it is not the particular
internal structure of the domains, but their spatial arrangement 
which determines the global conformations of the SCNPs (sparse Gaussian-like objects or crumpled globules).
Since the intramolecular stiffness can be tuned in real polymers
through the specific chemistry of the precursor or
by the introduction of bulky side groups, our results propose a new route to control
the global structure of SCNPs.

The article is organized as follows. In Section 2 we present details of the model and the simulation method. In Section 3 we characterize and discuss the effect of chain stiffness on the structure of the SCNPs and their domains.  Conclusions are given in Section 4.

\section{Model and simulation details}

We use a bead-spring model \cite{Kremer1990} to simulate the linear precursors and the SCNPs.
The non-bonded interactions between monomers are given by a purely repulsive Lennard-Jones (LJ) potential, 
\begin{equation}
U^{\rm LJ}(r) = 4\epsilon \left[ \left(\frac{\sigma}{r}\right)^{12} -\left(\frac{\sigma}{r}\right)^{6} +\frac{1}{4}\right] ,
\label{eq:LJ}
\end{equation}
with a cutoff distance $r_{\rm cut} = 2^{1/6}\sigma$. Connected monomers along the chain contour, as well as cross-linked monomers after synthesis of the SCNPs, interact via a finitely extensible nonlinear elastic (FENE) potential,
\begin{equation}
U^{\rm FENE}(r) = - \epsilon K_{\rm F} R_0^2 \ln \left[ 1 - \left(  \frac{r}{R_0}\right)^2 \right]  , 
\label{eq:fene}
\end{equation}
with $K_{\rm F} = 15$ and $R_0 = 1.5$. Chain stiffness is implemented through a bending potential,
\begin{equation}
U^{\rm bend}(\theta) = k\epsilon \left[ 1 - \cos\theta \right]  , 
\label{eq:bend}
\end{equation}
with $\theta$ the angle between consecutive bonds. In what follows we employ standard LJ units, $\epsilon = \sigma = m = 1$ (with $m$ being the monomer mass), setting the energy, length and time ($(\sigma^2m/\epsilon)^{1/2}$) scales, respectively.

A given fraction $f$ of the monomers in the precursor are reactive and can form irreversible bonds with other reactive monomers.
The reactive monomers are monofunctional, i.e., if two of them form a mutual bond this becomes permanent and 
they are not allowed to form new bonds. A bond between two unreacted monomers is formed when they are separated
by less than the `capture' distance $r < 1.3\sigma$. A random selection is made when there are several candidates within the capture distance. The reactive monomers are randomly distributed along the precursor
backbone, with the constraint that the placement of consecutive reactive monomers is forbidden, in order to prevent trivial cross-links.  

In all cases the precursors and the obtained SCNPs consist of $N=400$ monomers. The linear precursors
are first equilibrated without allowing the reactive monomers to form bonds. After equilibration cross-linking 
of the reactive monomers is initiated. Though the simulated polymers are coupled to the same thermal bath they are propagated
independently, i.e., cross-linking is purely intramolecular by construction.
When cross-linking is finished an acquisition run is performed for statistical averages.
In all the simulations the polymers are propagated under Langevin dynamics at fixed temperature $T = \epsilon/k_{\rm B}=1$.
We explore a broad range of fractions of reactive monomers $0.05 \leq f \leq 0.5$.
For each value of $f$ we perform simulations for bending constants $k = 0$ (fully-flexible case), 3, 5 and 8.
The characteristic ratio of the linear precursors is defined as the long-$N$ limit of $C_{\infty} = \langle R^2_{\rm e}\rangle/(Nb^2)$, with $\langle R^2_{\rm e}\rangle$ the end-to-end distance in the melt state, and $b \approx 0.95$
the mean bond length. For the fully-flexible case ($k=0$) a value $C_{\infty} =1.7$ is found \cite{Auhl2003}.
The calculation of $C_{\infty}$ for the investigated semiflexible chains by simulating the corresponding melts
is computationally expensive since convergence to the long-$N$ limit is much slower than for $k=0$.
However, an accurate estimation \cite{Auhl2003} can be obtained as 
$C_{\infty}= (1+ \langle \cos\theta \rangle)/(1- \langle \cos\theta \rangle)$, with
\begin{equation}
\langle \cos\theta \rangle =\frac{1+ \e^{2\beta k}(\beta k -1) +\beta k}{(\e^{2\beta k} -1)\beta k}
\label{eq:cinfty}
\end{equation}
and $\beta = (k_{\rm B}T)^{-1}$. By using the former equations we find $C_{\infty} \approx 5$, 9 and 15
for $k = 3$, 5 and 8, respectively. 
The investigated range $5 \lesssim C_{\infty} \lesssim 15$ for the semiflexible cases
corresponds to the majority of common polymers \cite{Rubinstein2003,Fetters2007}.
For each pair ($f,k$) we generate typically 100-200 realizations of the SCNPs.
Further details about the simulation method can be found in Ref.~\cite{Moreno2013}.

\section{Results and discussion}

\subsection{Size, shape and topological polydispersity}


\begin{figure}[htb!]
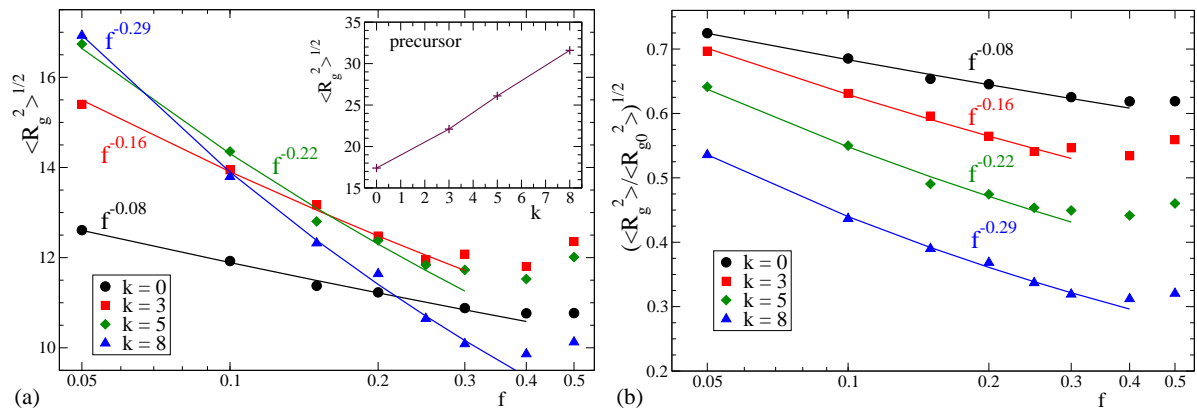

\centering
\includegraphics[width=.49\textwidth]{plot-f-rg-N400.eps}\hspace{0.2cm}\includegraphics[width=.49\textwidth]{plot-f-rgdrg0-N400.eps}
\caption{(a): Radius of gyration of the SCNPs as a function of the fraction $f$ of reactive monomers.
Different data sets correspond to different bending constants. The curves are fits to power-law behaviour
$\langle R_{\rm g}^2\rangle^{1/2} \sim f^{-z}$, with $z = 0.08, 0.16, 0.22$ and 0.29 
for $k = 0, 3, 5$ and 8,
respectively. The inset shows the $k$-dependence of the precursor size. (b): As panel (a), but after normalizing the SCNP sizes by those of the respective precursors.}
\label{fig:f-rg} 
\end{figure}

We first analyze the size and shape of the SCNPs.
Fig.~\ref{fig:f-rg} shows the average radius of gyration, $\langle R_{\rm g}^2\rangle^{1/2}$, 
of the SCNPs for all the investigated systems.
Increasing stiffness leads to a stronger dependence of the molecular size on the fraction of reactive monomers.
Not surprisingly, stiffness increases the size of the linear precursors with the same $N$ 
(by a factor 2 in the investigated range of $k$,
see inset in Fig.~\ref{fig:f-rg}a). However, the bending of 
segments of the semiflexible backbone into permanent loops that are, in average, longer in the SCNPs with higher $k$ 
(see next subsection) 
leads to a stronger relative shrinkage of the stiffest SCNPs by increasing $f$ (Fig.~\ref{fig:f-rg}b).
This can result in a non-monotonous $k$-dependence of the SCNP size at fixed $f$ 
(see data for $f \geq 0.1$ at Fig.~\ref{fig:f-rg}a).
The data for $f \leq 0.3$ can be described by a power-law $\langle R_{\rm g}^2\rangle^{1/2} \sim f^{-z}$, with $z = 0.08, 0.16, 0.22$ and 0.29 for $k = 0, 3, 5$ and 8, respectively.
It is worthy of remark that the exponent $z = 0.22$ found for $k=5$ ($C_{\infty} \approx 9$)
is consistent with that observed for SCNPs based in polystyrene \cite{DelaCuesta2017}, which has a similar $C_{\infty} \approx 10$ \cite{Rubinstein2003}. As observed in simulations of polystyrene-based SCNPs \cite{Liu2009}, 
increasing the fraction of monomers above 30\% does not further result in
a significant reduction of the molecular size, and even a slight swelling effect is found. 
It must be stressed that, unlike in the simulations of Ref.~\cite{Liu2009}, cross-linking of the SCNPs 
has been completed (i.e., all the reactive monomers have formed bonds). 
Therefore we conclude that for $f > 0.3$ adding more reactive monomers
just contributes to the formation of small loops, which are inefficient for further compaction of the SCNPs.

\begin{figure}[htb!]
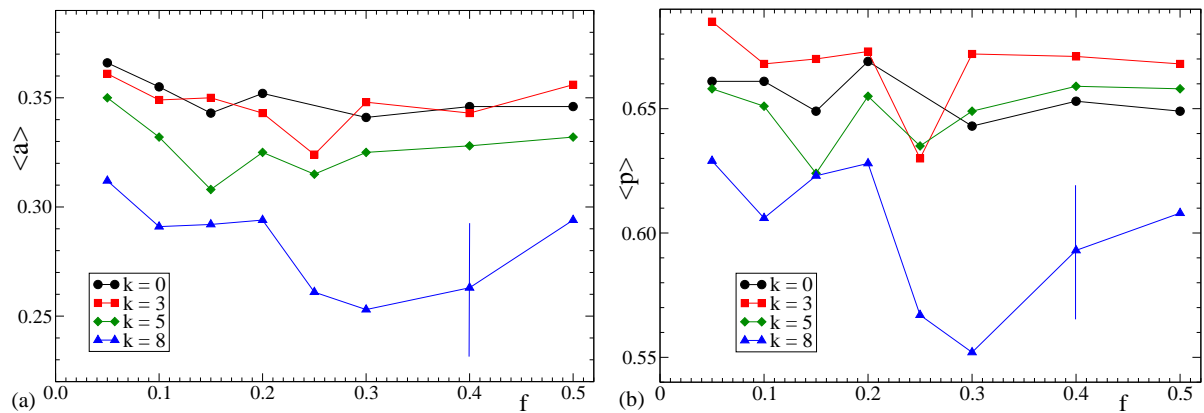

\centering
\includegraphics[width=.49\textwidth]{plot-f-averasph-N400.eps}\hspace{0.3cm}\includegraphics[width=.49\textwidth]{plot-f-averprol-N400.eps}
\caption{Mean asphericity (a) and prolateness (b) of the SCNPs as a function of the fraction of reactive monomers.
Different data sets correspond to different values of the bending constant. The vertical lines are typical error bars.}
\label{fig:f-asph-prol} 
\end{figure}

Insight on the global shape of the SCNPs can be obtained by
computing the asphericity $a$ and prolateness $p$ parameters \cite{rawdon2008}. These are defined as:
\begin{equation}
a = \left\langle \frac{(\lambda_2 -\lambda_1)^2 +(\lambda_3 -\lambda_1)^2 +(\lambda_3 -\lambda_2)^2}{2(\lambda_1 +\lambda_2 +\lambda_3)^2} \right\rangle ,
\end{equation}
\begin{equation}
p = \left\langle \frac{(3\lambda_1 -R^2_{\rm g})(3\lambda_2 -R^2_{\rm g})(3\lambda_3 -R^2_{\rm g}) }{2(\lambda_1^2 +\lambda_2^2 +\lambda_3^2 -\lambda_1\lambda_2 -\lambda_1\lambda_3 -\lambda_2\lambda_3)^{3/2}} \right\rangle ,
\end{equation}
where $\lambda_1, \lambda_2 , \lambda_3$ are the eigenvalues of the radius of gyration tensor.
The asphericity $0 \leq a \leq 1$ quantifies deviations from spherosymmetrical shape ($a=0$). 
The prolateness varies between the limits of perfectly oblate ($p=-1$) and prolate ($p=1$) objects. 
Fig.~\ref{fig:f-asph-prol} shows the mean asphericity and prolateness for all the investigated systems.
No clear trend is found in the global shape of the SCNPs by increasing the fraction of reactive monomers.
Instead, increasing stiffness at fixed $f$ leads to SCNPs that are, in average, more spherical and less prolate. 
As will be shown later, this is consistent with the increasing population of long loops at higher $k$, 
which leads to a more efficient compaction of the SCNPs and will be reflected in their scaling behaviour.

\begin{figure}[htb!]
\centering
\includegraphics[width=.49\textwidth]{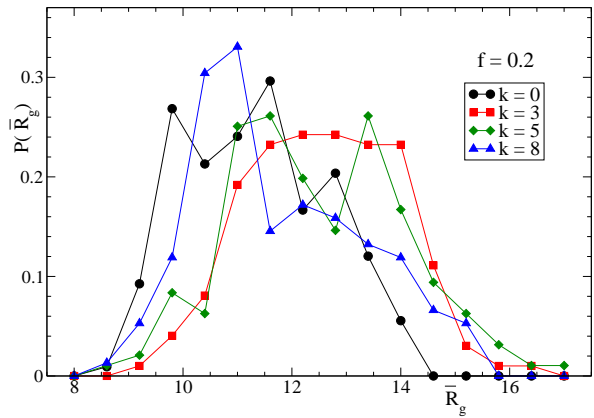}
\caption{Distributions of the time-averaged radius of gyration in SCNPs with $f = 0.2$.
Different data sets correspond to different bending constants.}
\label{fig:dist-rg} 
\end{figure}
\begin{figure}[htb!]
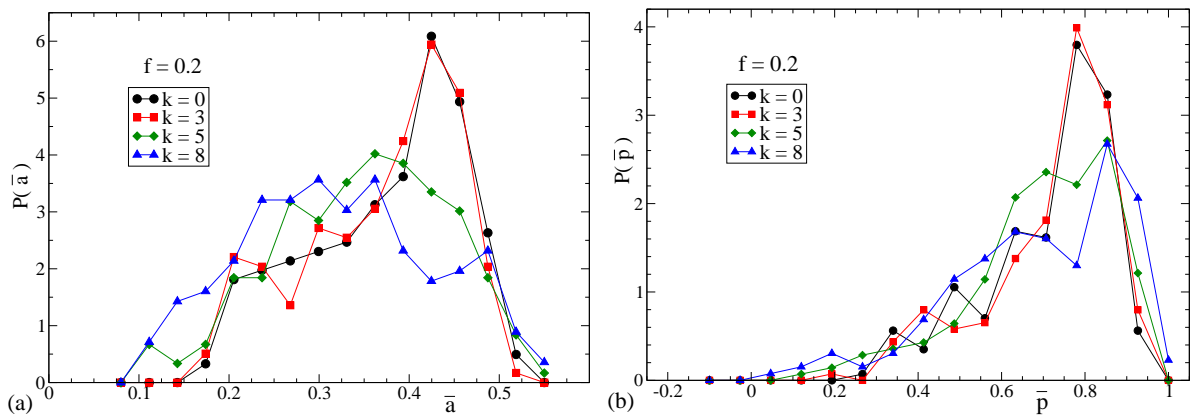

\centering
\includegraphics[width=.49\textwidth]{plot-timav-asph-N400L80.eps}\hspace{0.2cm}\includegraphics[width=.49\textwidth]{plot-timav-prol-N400L80.eps}
\caption{Distributions of the time-averaged asphericity (a) and prolateness (b) in SCNPs with $f = 0.2$.
Different data sets correspond to different bending constants.}
\label{fig:dist-asph-prol} 
\end{figure}

As aforementioned, SCNPs are topologically polydisperse objects even if they are sythesized from the same precursor (same chemistry, molecular weight
and fraction of reactive groups.)
To account for the effect of stiffness in the topological polydispersity we have computed for each individual
SCNP its time-averaged size and shape parameters (i.e., averaged over its internal fluctuations along the simulation trajectory).
We denote these time averages as $\bar{R_{\rm g}}$, $\bar{a}$ and $\bar{p}$. 
Figs.~\ref{fig:dist-rg} and \ref{fig:dist-asph-prol}
show representative results for the distributions of such parameters at fixed $f = 0.2$ and all the investigated bending constants. 
In spite of the poor statistics, the distributions at fixed $f$ seem to shift with increasing $k$ following the same trends that the mean values in Figs.~\ref{fig:f-rg}a and \ref{fig:f-asph-prol} (non-monotonous and monotonous 
for the size and shape parameters, respectively). No clear effect is found in the width of the distributions, though increasing stiffness seems to produce more symmetrical distributions of the shape parameters.

\subsection{Connectivity and scaling}

\begin{figure}[htb!]
\centering
\small
(a)\includegraphics[width=.43\textwidth]{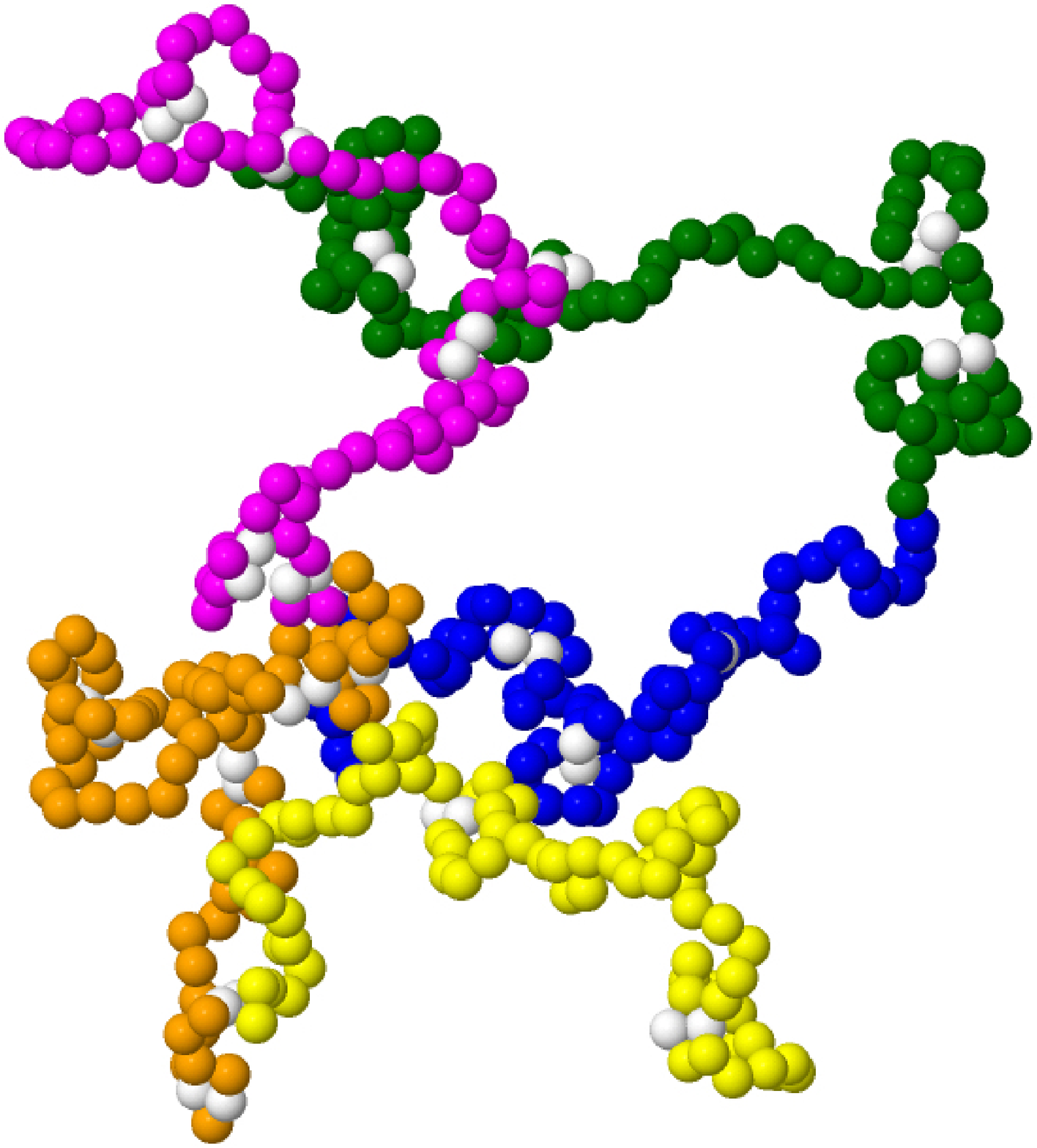}\hspace{0.2cm}(b)\includegraphics[width=.43\textwidth]{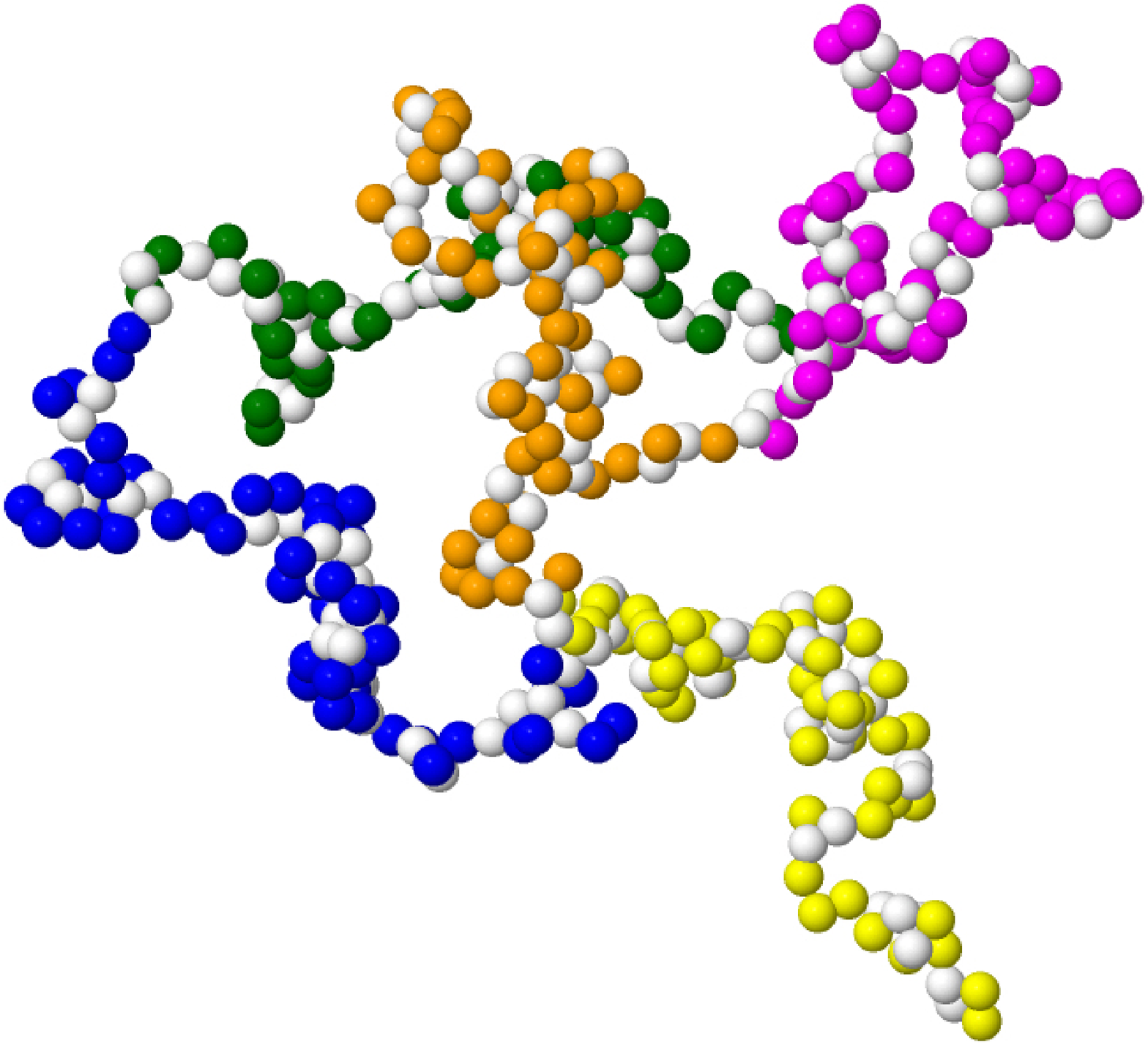}
(c)\includegraphics[width=.43\textwidth]{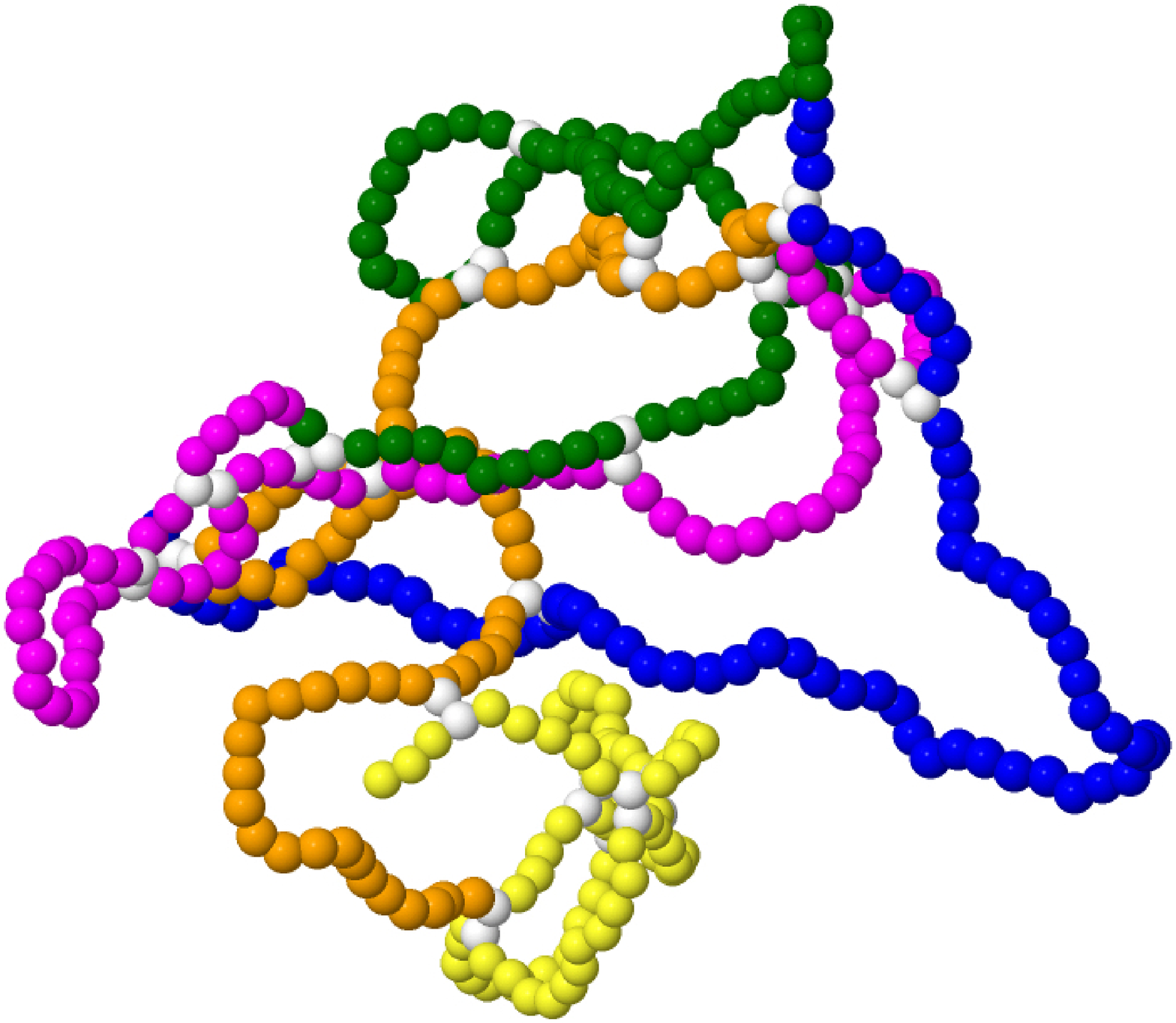}\hspace{0.2cm}(d)\includegraphics[width=.45\textwidth]{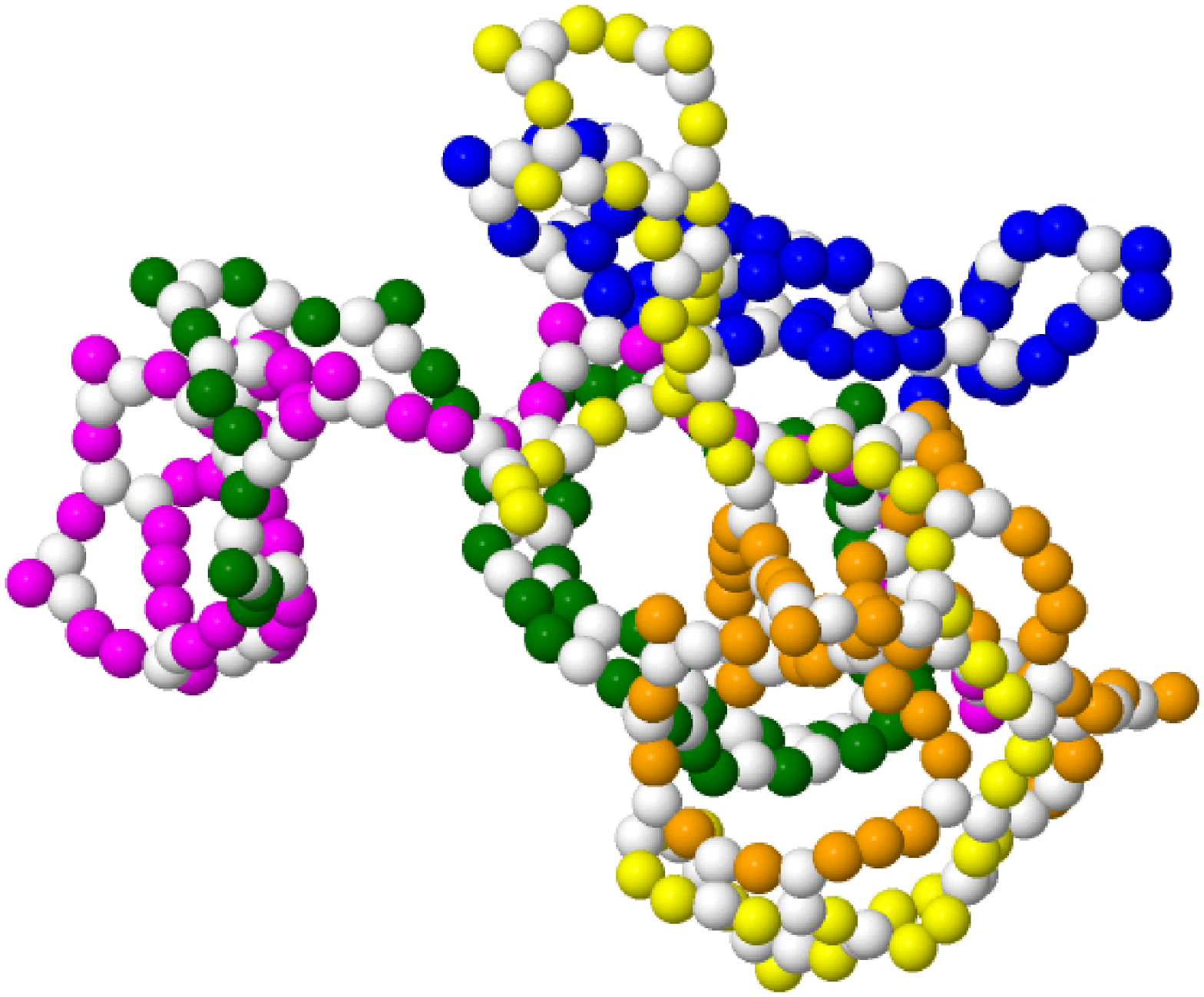}
\normalsize
\caption{Typical SCNP conformations with different values of the bending constant $k$ and fraction of reactive monomers $f$. 
(a): $k=0, f=0.1$, (b): $k=0, f=0.4$, (c): $k=8, f=0.1$, (d): $k=8, f=0.4$. Cross-linked monomers are depicted in white.
The rest of the monomers are depicted according to their positions $1 \le i \le N=400$ in the linear precursor backbone.
Yellow: $1 \le i \le 80$; orange: $81 \le i \le 160$; magenta: $161 \le i \le 240$; green: $241 \le i \le 320$; 
blue: $321 \le i \le 400$. }
\label{fig:snapcols} 
\end{figure}

Fig.~\ref{fig:snapcols} shows some representative snapshots of SCNPs in the fully-flexible ($k=0$) and the stiffest investigated case ($k=8$), in both cases with two different fractions of reactive monomers $f = 0.1$ and $f=0.4$. The cross-linked reactive monomers are depicted in white. The rest of the monomers are depicted in colours, following the rainbow sequence, according to their positions in the linear precursor. 
As can be recognized by the high presence of cross-linked white monomers within segments of the same colour,
short and middle-range loops largely dominate the SCNP connectivity. Only
a few long loops (cross-links between segments far in the rainbow sequence) are present.

\begin{figure}[htb!]
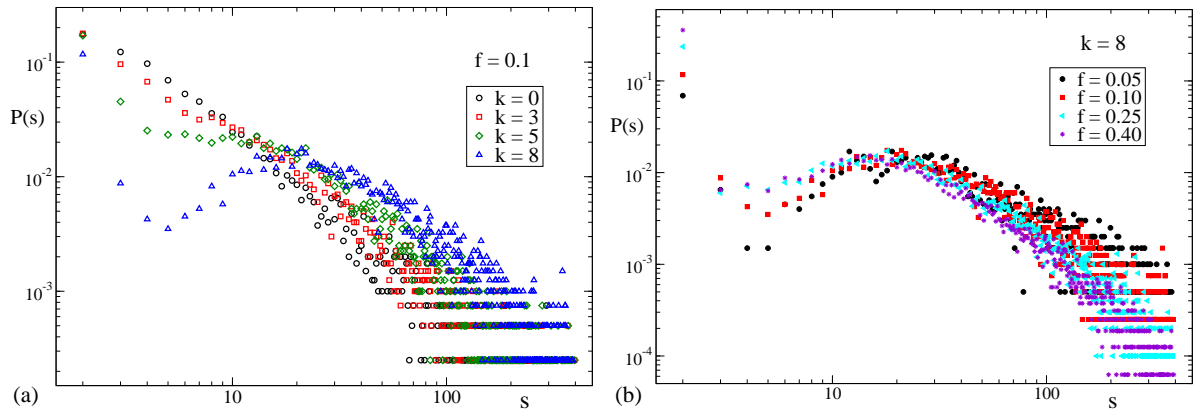

\centering
\includegraphics[width=.49\textwidth]{plot-histog-N400-L40.eps}\hspace{0.2cm}\includegraphics[width=.49\textwidth]{plot-histog-N400-k8-comp-f.eps}
\caption{Distribution of contour distances between bonded reactive monomers. (a): data at fixed $f=0.1$ for different bending
constants. (b): data at fixed $k=8$ for different fractions of reactive monomers.}
\label{fig:histog} 
\end{figure}

By labelling the monomers as $i=1,2,...N$ from one end of the precursor backbone to the other one,
we define the contour distance between two bonded reactive monomers $i,j$ as $s = |i-j|$.
A characterization of the effect of chain stiffness and fraction of reactive monomers 
on the connectivity of the SCNPs can be obtained by representing the distribution $P(s)$, as shown in Fig.~\ref{fig:histog}.
Panel (a) shows results for different bending constants $k$ at the fixed fraction of reactive monomers $f = 0.1$.
Chain stiffness has a dramatic effect on the shape of the distributions. In all cases $P(s)$ shows within statistics a monotonous decay  for long distances. This is a consequence of the self-avoiding character of the precursor in the
good solvent conditions of synthesis, which makes contacts at long contour distances, 
and the corresponding cross-linking events, unfrequent.
The value of the bending constant has a very different impact at short and intermediate distances.
For the two lowest values $k= 0$ and 3 the monotonous decay is observed in the whole
range of contour distances.  
A well-defined plateau  arises for $k=5$ in the range $4 \lesssim s \lesssim 20$.
For $k =8$, $P(s)$ shows a complex behavior, with a steep decay to a local minimum at $s = 5$, followed by an increase
and a plateau at $10 \lesssim s \lesssim 40$ prior the final decay at long $s$.
In none of the investigated cases stiffness is strong enough to prevent bonds between the closest reactive monomers
(forming triangle and square loops for $s=2$ and $s=3$, respectively), which are still dominant in the distribution.
However, loops with $s>3$ but still shorter than one Kuhn step \cite{Rubinstein2003} 
(i.e., $s < C_{\infty} \approx 15$ monomers) are disfavoured for the stiffest investigated systems ($k=8$).
This kind of loops indeed involve approximately antiparallel alignments of neighbouring strands 
(see bottom panels in Fig.~\ref{fig:snapcols}) with strong 
bending penalties at the turning points. 
Obviously, in the fully cross-linked SCNPs with identical $N$ and fraction of cross-links, 
a lower population of small loops 
involves a higher population of long-range loops, as it is found in $P(s)$ by increasing $k$.
Fig.~\ref{fig:histog}b shows the distributions $P(s)$ for fixed $k = 8$ as a function of $f$.
Increasing $f$ has the opposite effect to increasing $k$. It leads to shorter contour distances between consecutive reactive monomers, which favours bonding at short $s$ and partially compensates the disfavouring effect of chain stiffness.

\begin{figure}[htb!]
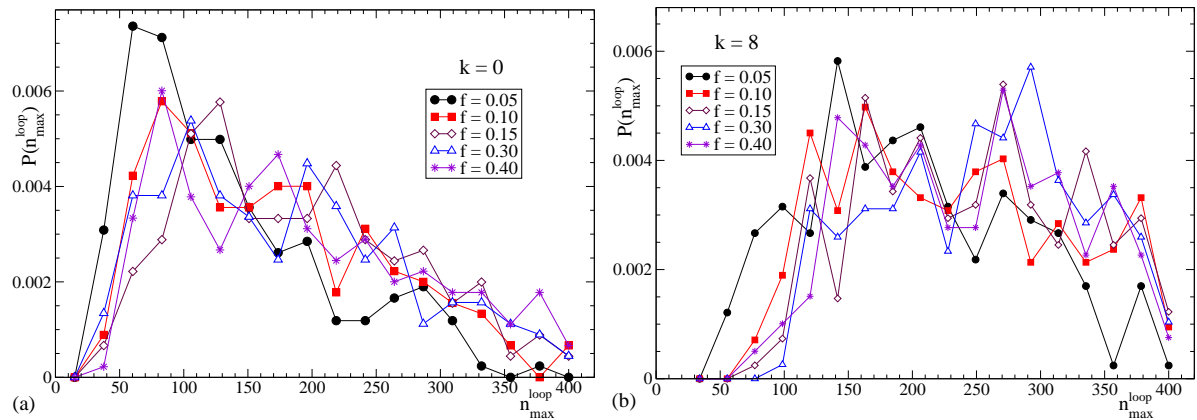

\centering
\includegraphics[width=.49\textwidth]{plot-maxloop-f-k0.eps}\hspace{0.2cm}\includegraphics[width=.49\textwidth]{plot-maxloop-f-k8.eps}
\caption{Distribution of the biggest loop in SCNPs with $k=0$ (a) and $k=8$ (b).
Different data sets correspond to different values of the fraction of reactive monomers.}
\label{fig:maxloop} 
\end{figure}

Though, as revealed by the distributions $P(s)$, long-range loops are unfrequent, most of the SCNPs have at least one big loop
of $N > 100$ monomers. This is demonstrated in Fig.~\ref{fig:maxloop}. For each individual SCNP we identify its biggest loop,
of $n^{\rm loop}_{\rm max}$ monomers. Panels (a) and (b) show the corresponding distributions $P(n^{\rm loop}_{\rm max})$ 
for $k=0$ and $k=8$, respectively, and at different values of $f$. Within the poor statistics there is no apparent effect 
of $f$ (provided that $f \ge 0.10$) in the size of the biggest loop. On the contrary, increasing stiffness has a clear effect,
leading to higher values of the mean biggest loop. For $k=8$ the distributions  $P(n^{\rm loop}_{\rm max})$
become symmetric. For $k=0$ (panel (a)) as well as for $k=3$  and 5 (not shown), the distributions are asymmetric
with a maximum in the low-$n^{\rm loop}_{\rm max}$ range. This suggests that large-scale conformations of SCNPs with $k \le 5$ are more 
`chain-like', whereas those for $k=8$ are more `ring'-like. Indeed, for $k=8$ the biggest loop in the SCNP
typically contains more than 60\% of the monomers.

\begin{figure}[htb!]
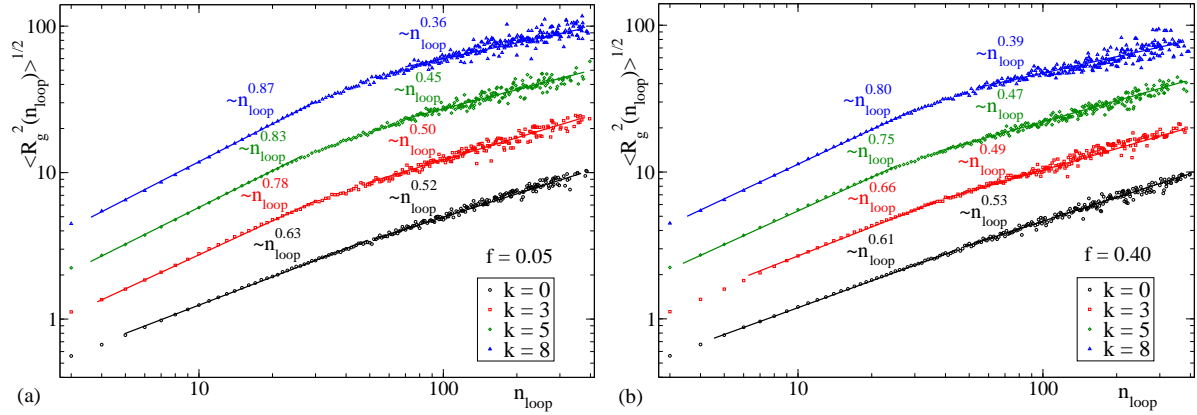

\centering
\includegraphics[width=.49\textwidth]{plot-rgloop-N400L40.eps}\hspace{0.2cm}\includegraphics[width=.49\textwidth]{plot-rgloop-N400L160.eps}
\caption{Radius of gyration of the loops as a function of their number of monomers. 
(a): data for $f=0.05$; (b): data for $f=0.40$. Different data sets correspond to different bending constants.
For the sake of clarity each set has been rescaled in the ordinate axis, from bottom to top,
by a factor 1, 2, 4 and 8. 
Lines are fits to power-law behaviour, $\langle R^2_{\rm g}(n_{\rm d}) \rangle  \sim n_{\rm d}^{\mu'}, n_{\rm d}^{\mu}$,
with $\mu'$ and $\mu$ the exponents for small and long loops. Their values are indicated.}
\label{fig:rgloop} 
\end{figure}

Increasing stiffness does not only have a dramatic effect on the distribution $P(s)$ of loop sizes (Fig.~\ref{fig:histog}), but also on the shape of the formed loops. This is demonstrated by representing the radius of gyration 
of the loop vs. its number of monomers $n_{\rm loop}$.
Results are shown in Fig~\ref{fig:rgloop} as a function of $k$ for two fractions of reactive groups $f = 0.05$ and $f=0.40$. 
The data are consistent with power-law behaviour $\langle R^2(s)\rangle^{1/2} \sim n_{\rm loop}^{\mu}$, with different exponents 
for small and long loops. 
Not surprisingly, for small loops ($s < 40$) the exponent $\mu$ increases with the bending constant, reflecting the increasing persistence length. 
The exponent takes values that are between the limits of self-avoiding 
random walks ($\mu \approx 0.59$) \cite{Rubinstein2003} and rods ($\mu =1$).
The power-laws for long loops are characterized by lower $\mu$-exponents,
with no significant dependence of $\mu$ on the fraction of reactive monomers.
Increasing stiffness has the opposite effect to that observed at low $n_{\rm loop}$, resulting in decreasing exponents, which suggests
increasing compactness of the long loops.
The exponents change from $\mu \approx 0.5$ for fully-flexible SCNPs ($k = 0$) to $\mu \lesssim 0.4$ for 
the stiffest investigated ones ($k= 8$).

\begin{figure}[htb!]
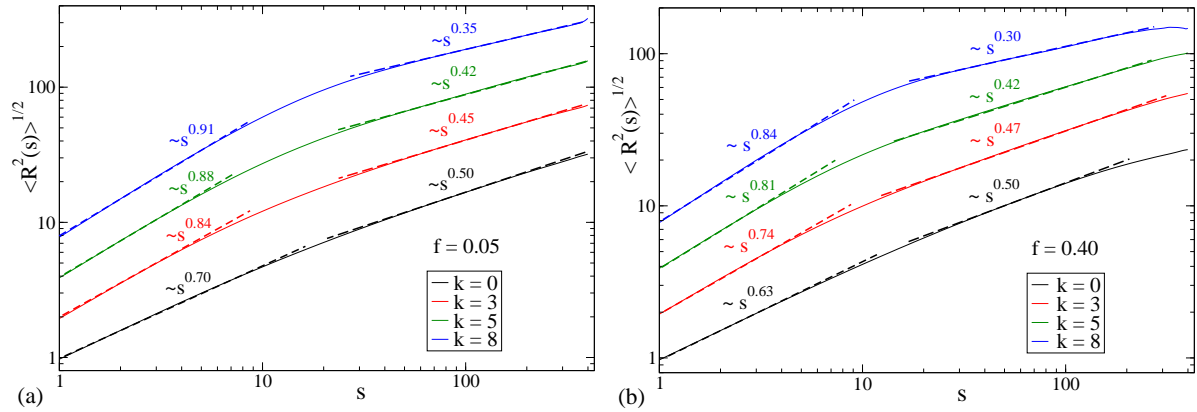

\centering
\includegraphics[width=.49\textwidth]{plot-rs-N400L20.eps}\hspace{0.2cm}\includegraphics[width=.49\textwidth]{plot-rs-N400L160.eps}
\caption{Real-space distance vs. contour distance for SCNPs with fixed $f=0.05$ (a) and $f=0.40$ (b).
Different data sets correspond to different bending constants. For the sake of clarity each set has been rescaled in the ordinate axis, from bottom to top, by a factor 1, 2, 4 and 8. Solid lines are simulation data. 
Dashed lines are fits to power-laws $\langle R^2(s)\rangle^{1/2} \sim s^{p'}, s^p$, with $p'$ and $p$
the exponents for short and long distances. Their values are indicated.}
\label{fig:rs} 
\end{figure}

Analogous results are found for the average real space distance, $\langle R^2(s)\rangle^{1/2}$, 
between {\it any two} monomers in the SCNP separated by a contour distance $s$.
Fig.~\ref{fig:rs} shows results for $\langle R^2(s)\rangle^{1/2}$ at the same values of $(f,k)$ as in Fig.~\ref{fig:rgloop}.
The data are again consistent with power-law behaviour, $\langle R^2(s)\rangle^{1/2} \sim s^p$, 
with different exponents at short and long distances. 
Again for short paths ($s < 10$) the exponent increases with the bending constant, reflecting the increasing persistence length,
and varies between the self-avoiding chain and rod-like limits.
Increasing the degree of cross-linking leads to more return events, resulting in a weaker dependence
of the real-space distance on the contour distance and therefore slightly lower exponents.
The power-laws at long contour distances ($s > 20$) are characterized by lower $p$-exponents than at short $s$ and,
as in Fig.~\ref{fig:rgloop}, with no significant dependence of $p$ on the fraction of reactive monomers.
Again increasing stiffness has the opposite effect in the exponents that at short paths, highlighting the increasing compactness
of the global conformations. In a similar fashion to Fig.~\ref{fig:rgloop},
the exponents change from $p \approx 0.5$ for $k = 0$ to $p \approx 0.3$ for $k= 8$.

\begin{figure}[htb!]
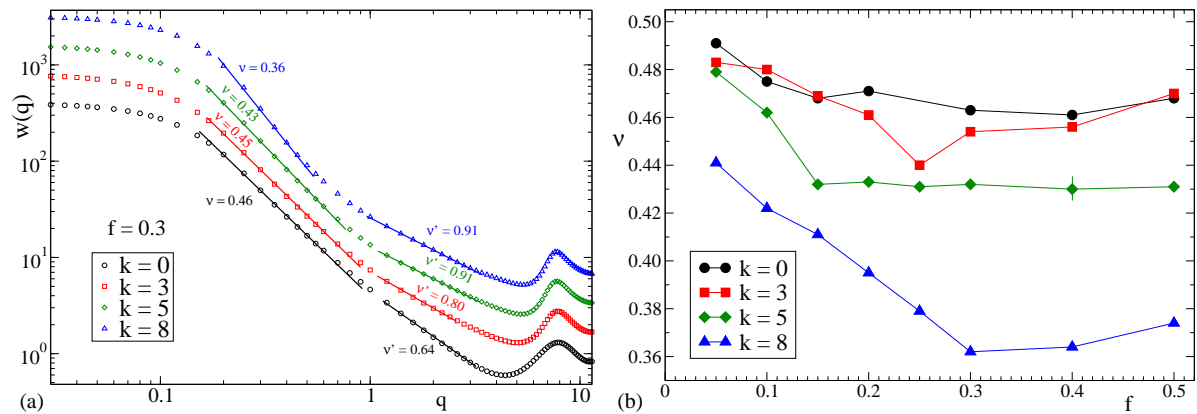

\centering
\includegraphics[width=.49\textwidth]{plot-wq-N400L120.eps}\hspace{0.2cm}\includegraphics[width=.49\textwidth]{plot-f-nu.eps}
\caption{(a): Form factors for SCNPs with $f=0.3$. Different data sets correspond to different bending constants.
For the sake of clarity each set has been rescaled in the ordinate axis, from bottom to top, by a factor 1, 2, 4 and 8.
Lines are power law fits: $w(q) \sim q^{-1/\nu}$ in the fractal regime $1/R_{\rm g} \ll q \ll 1/b$, 
and $w(q) \sim q^{-1/\nu'}$ at $1 < q < 4$. Exponents are indicated.
(b): Scaling exponents $\nu$ for the fractal regime as a function of the fraction of reactive monomers.
Different data sets correspond to different values of the bending constant.
A typical error bar is shown.}
\label{fig:wq} 
\end{figure}

Information about the intramolecular collective correlations can be obtained by analyzing the form factors.
These are calculated as
\begin{equation}
w(q) = \left\langle \frac{1}{N} \sum_{j,k} \exp\left[ i \mathbf{q}\cdot ( \mathbf{r}_j - \mathbf{r}_k )\right] \right\rangle \, , 
\label{eq:formfactor}
\end{equation}
where $\bf{q}$ is the wave vector and the sum is restricted over monomers belonging to the same SCNP. 
Fig.~\ref{fig:wq}a shows results for the form factors of SCNPs with fixed $f =0.3$ at all the investigated bending constants.
In the fractal regime, $1/R_{\rm g} \lesssim q \lesssim 1/b $, the form factor scales as 
$w(q) \sim q^{-1/\nu}$, with $\nu$ the scaling exponent \cite{Rubinstein2003}.
Fits to the former power-law are shown in Fig.~\ref{fig:wq}a. The obtained exponents for all the investigated systems
are displayed in Fig.~\ref{fig:wq}b. The exponents found in each system are similar to those observed in
the corresponding long-loop and long-distance scaling of $\langle R_{\rm g}^2(n_{\rm loop})\rangle^{1/2}$ 
and $\langle R^2(s)\rangle^{1/2}$, respectively, following the same trends
and confirming that increasing stiffness
leads to more compact SCNPs. The form factors also exhibit power-law behavior at $1 < q < 4$, corresponding to distances ($2\pi/q$) of about 2 to 6 monomer diameters. This $q$-regime probes the local self-avoiding structure of the backbone at small distances, and indeed
the effective exponents are consistent with those found for $\langle R_{\rm g}^2(n_{\rm loop})\rangle^{1/2}$
at small loops (Fig.~\ref{fig:rgloop}) and $\langle R^2(s)\rangle^{1/2}$ at short contour distances (Fig.~\ref{fig:rs}).

The exponents $\mu, p, \nu \lesssim 0.5$ found for $k \le 5$ 
suggest a structural analogy with Gaussian chains and rings in $\theta$-solvent,
where local globules are formed (produced by the short-range loops) but the global conformation is an open object 
(in general with at least one long loop,  as shown in Fig~\ref{fig:maxloop}).
The scaling with exponents $\mu, p, \nu \gtrsim 0.3$ found for the stiffest SCNPs ($k=8$) instead resembles that of a fractal or `crumpled' globule.
This is very different from the scaling expected for `equilibrium' globules, as it is the case of collapsed chains. In that case 
$\langle R^2(s)\rangle^{1/2}$ should obey Gaussian statistics as chains in a melt ($\sim s^{1/2}$), reaching a plateau at $s \approx (3N/4\pi)^{2/3} \approx 20$
where the radius of the confining sphere is reached and the Gaussian paths are bounced back \cite{Grosberg1988,Mirny2011}.  
Likewise, a dense equilibrium globule would lead to Porod scattering \cite{Rubinstein2003} $w(q) \sim q^{-4}$ in the form factor, i.e, an effective exponent $\nu = 0.25$ much lower than those observed for $k=8$ (Fig.~\ref{fig:wq}). 
In spite of these analogies, it must be noted that these refer to the large-scale properties of the SCNP, and that there
are still some differences with the crumpled globular structure. First, the approximate scaling $\langle R^2(s)\rangle^{1/2} \sim s^{1/3}$ found for $k=8$
breaks down below local but relatively large scales ($s \approx 20$, see Fig.~\ref{fig:rs}). Indeed there are no really dense `melt-like' 
regions in the stiff SCNPs (see bottom panels in Fig.~\ref{fig:snapcols}) resembling the `territories' 
observed in melts of rings and chromatime  \cite{Mirny2011,Halverson2011,Halverson2014}. These would indeed involve strong folding 
of chain segments and consequently a large bending penalty.

\begin{figure}[htb!]
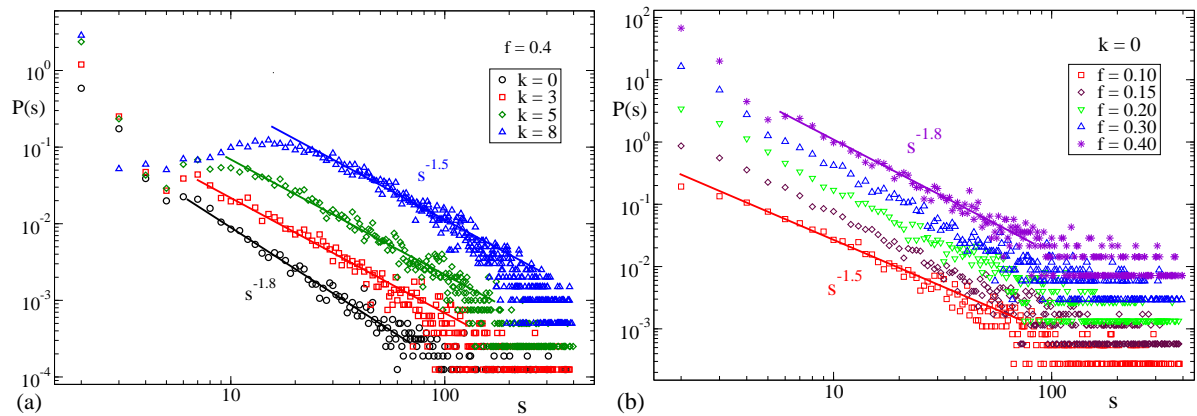

\centering
\includegraphics[width=.49\textwidth]{plot-histog-N400-L160-scal.eps}\hspace{0.2cm}\includegraphics[width=.49\textwidth]{plot-histog-k0-f-comp-scal.eps}
\caption{Distribution of contour distances between bonded reactive monomers.
(a): data at fixed $f=0.4$ for different bending
constants. (b): data at fixed $k=0$ for different fractions of reactive monomers.
For the sake of clarity each set has been rescaled in the ordinate axis, from bottom to top,
by a factor 1, 3, 9, 27 and 81. The solid lines indicate approximate power law behavior $P(s) \sim s^{-x}$ with
$1.5 \leq x \leq 1.8$. }
\label{fig:histogscal} 
\end{figure}

The second, and more relevant, difference is that the SCNPs do not show the 
approximate scaling $P(s) \sim s^{-1}$ expected for crumpled globules \cite{Mirny2011,Halverson2011,Halverson2014,Rosa2010}.
The decay of $P(s)$ at long contour distances is, for all the investigated values of $k$ and $f$, 
compatible with power-law behaviour, $P(s) \sim s^{-x}$, but with an exponent $1.5 \le x \le 1.8$
(see results for $f=0.4$ as a function of $k$ and for $k=0$ as a function of $f$ in both panels of Fig.~\ref{fig:histogscal}).
Similar exponents have been found in a similar on-lattice model for SCNPs \cite{Rabbel2017}.
For the flexible systems the exponent $x \approx 1.5$ can be rationalized by invoking Gaussian statistics of linear chains, 
for which the return probability of a long segment ($s \gg 1$) scales as $s^{-3/2}$ \cite{Lua2004,notedefps}.

\begin{figure}[htb!]
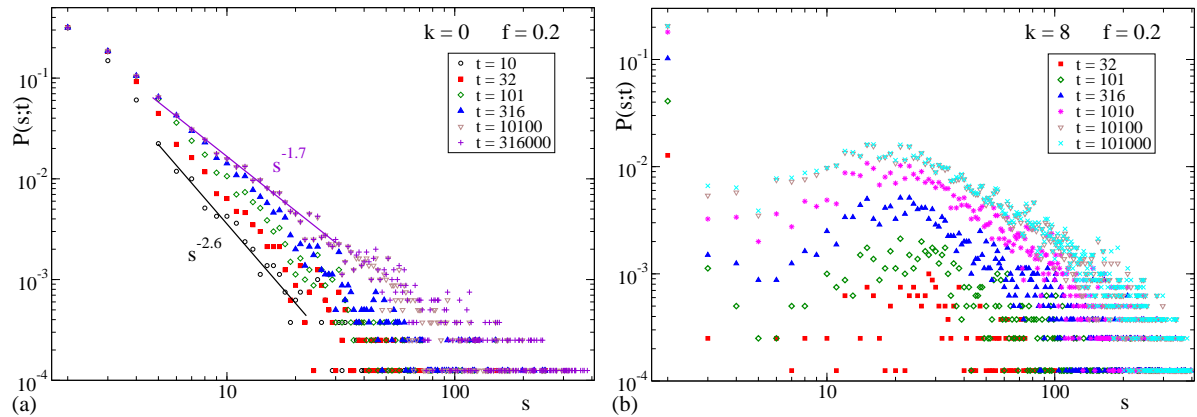

\centering
\includegraphics[width=.49\textwidth]{plot-histogtime-N400L80-k0.eps}\hspace{0.2cm}\includegraphics[width=.49\textwidth]{plot-histogtime-N400L80-k8.eps}
\caption{Unnormalized distributions (see text) of contour distances between bonded reactive monomers in SCNPs with $f=0.2$.
Panels (a) and (b) correspond to bending constants $k=0$ and $k=8$, respectively.
Different data sets correspond to different times during the cross-linking run.
Lines in panel (a) are fits to power-law behaviour $P(s;t) \sim s^{-x}$, with $x = 2.6$ and 1.7 at early and late times, respectively.}
\label{fig:histogtime} 
\end{figure}


At this point there is an apparent contradiction between the scaling properties found for $k=8$
in the loop size, the internal distance and the form factor (consistent with a crumpled globular
structure) and the approximate behaviour $P(s) \sim s^{-x}$ with $x \gtrsim 1.5$ instead of $x \approx 1$
(and hence consistent with Gaussian or self-avoiding chains). 
Though we do not have a strong argument for solving this apparent contradiction, a temptative
explanation can be obtained by analyzing the kinetics of cross-linking. 
Fig.~\ref{fig:histogtime} shows, for fixed $f=0.2$ and two bending constants
$k=0$ and $k=8$, the time evolution of the distributions $P(s)$ 
during the cross-linking process. For a correct comparison between distributions at different times,
the total number of counts at each time is divided by the total number of SCNPs and the maximum number of bonds (e.g., 40 for $f=0.2$). In this way $\int P(s;t)ds = 1$ only at late times when cross-linking is completed. 
If cross-linking is not completed $\int P(s;t)ds $ accounts for the fraction of reactive monomers
that have formed a permanent bond at time $t$.
By comparing both panels of Fig.~\ref{fig:histogtime} we observe that the nature of the
cross-linking process is strongly affected by increasing stiffness.  
For the fully-flexible case ($k=0$) cross-linking at early times is largely dominated by the formation
of small loops. These short-range events saturate at intermediate times and long loops are formed significantly only at late times (note the decreasing slope  of $P(s;t)$ with time).
Since the relative fraction of middle and long loops is very small the SCNPs are, in average, sparse objects with just local globulation, 
resembling linear chains or rings in $\theta$-solvent, which rationalizes the approximate scaling $P(s) \sim s^{-3/2}$.
In the case of very stiff SCNPs ($k=8$), short- and long-range loops are formed roughly in a simultaneous fashion, without
a clear time scale separation. This can be seen in Fig.~\ref{fig:histogtime}b,
where the distributions $P(s;t)$ shift vertically with time but their shape is not significantly altered.
Therefore, a high fraction of the middle and long loops are already formed at early and intermediate times, where the still
weakly cross-linked precursors are open self-avoiding objects. In these conditions the probability
of forming a middle or long loop is roughly $P(s) \sim s^{-3/2}$. Compaction occurs at later times,
much of it being driven by cross-linking of big loops mediated by bonds over shorter contour distances.
This late process has only a minor effect in the total distribution $P(s)$,
which retains the scaling behaviour reached at intermediate times.   

\subsection{Characterization of SCNP domains}

\begin{figure}[htb!]
\centering
\small
(a)\includegraphics[width=.45\textwidth]{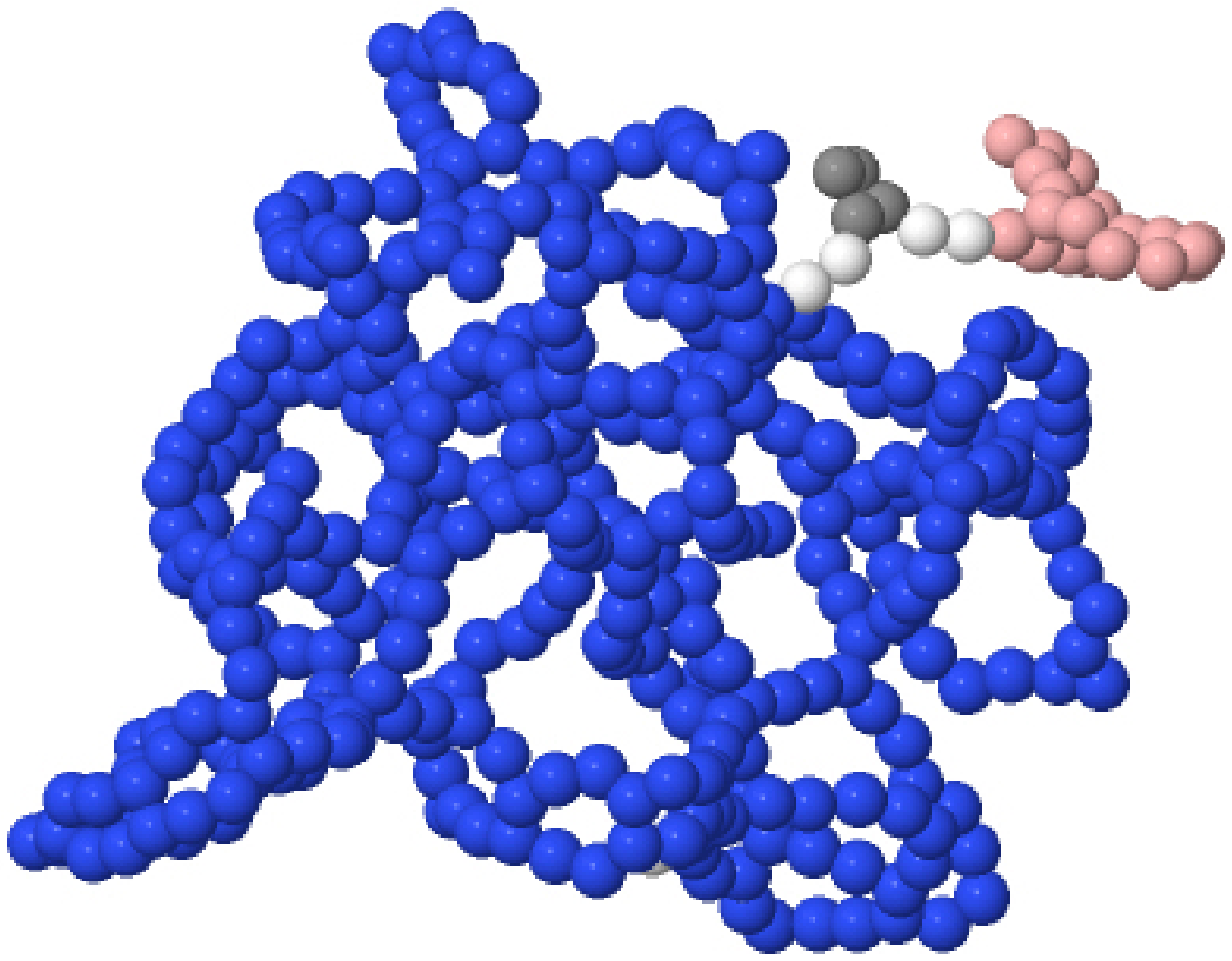}\hspace{0.2cm}(b)\includegraphics[width=.45\textwidth]{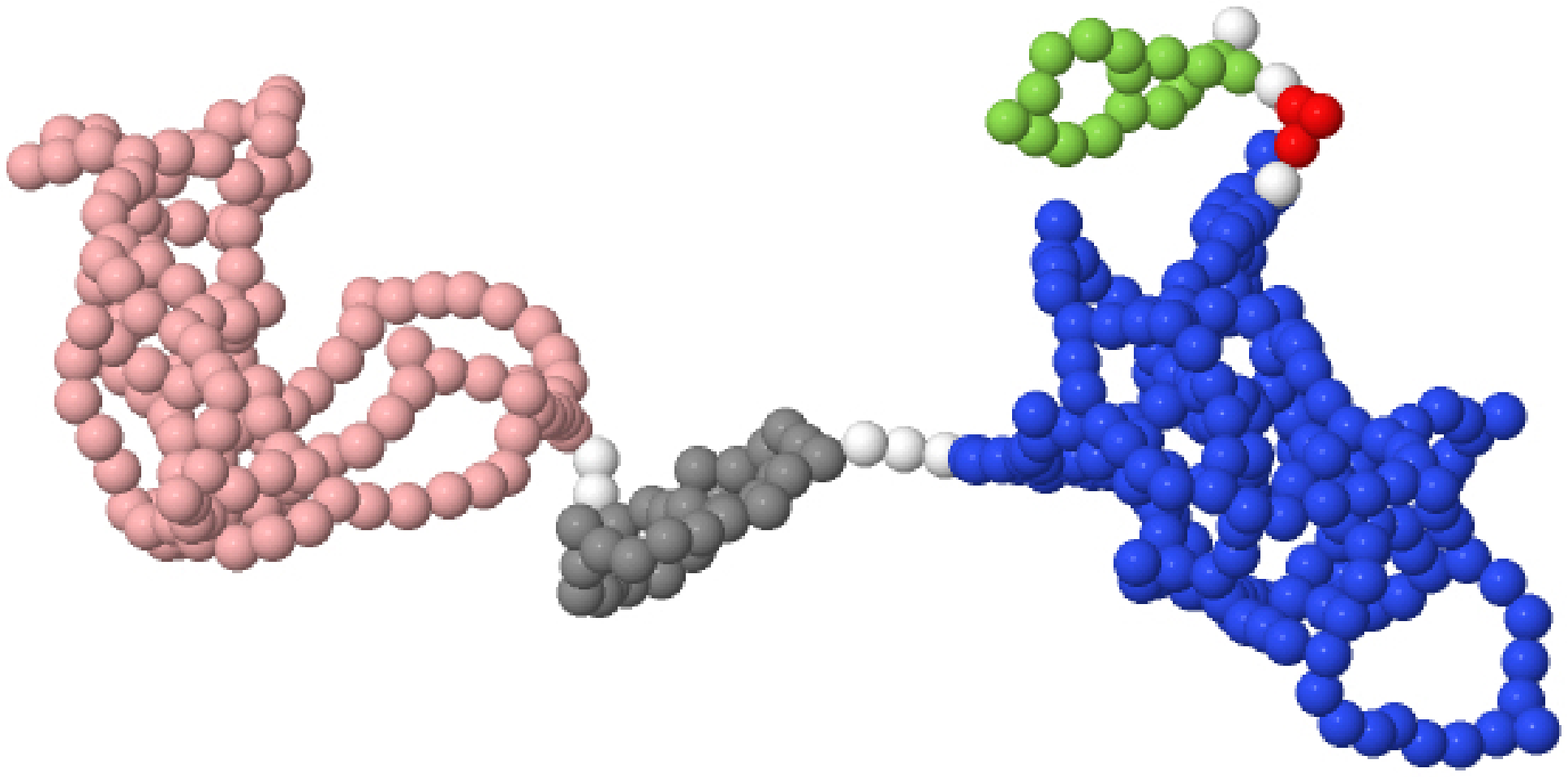}
\normalsize
\caption{Snapshots of two typical SCNPs with $k = 8$ and $f = 0.3$. Different domains are depicted in different colours. 
Monomers not belonging to domains are depicted in white. Panels (a) and (b) correspond to a SCNP
with low and high asphericity, respectively.}
\label{fig:snap-domains} 
\end{figure}

\begin{figure}[htb!]
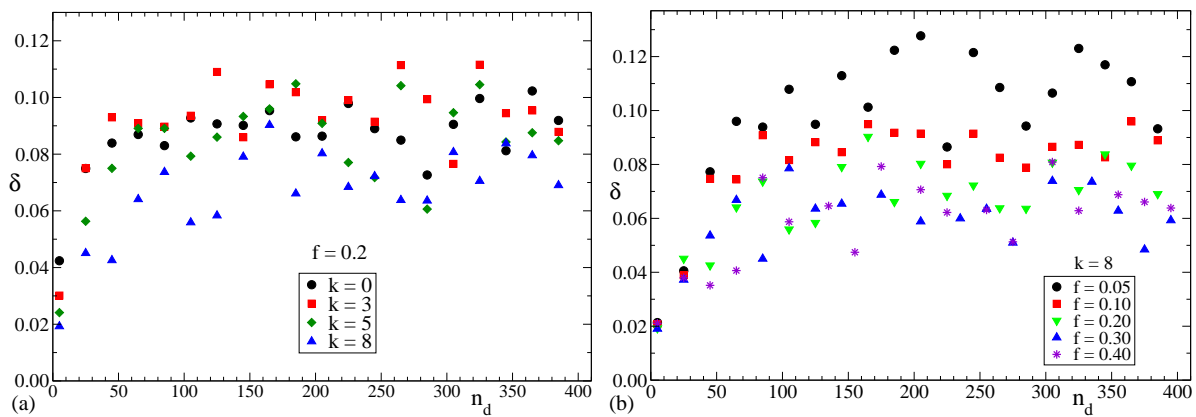

\centering
\includegraphics[width=.49\textwidth]{plot-fluctdomain-N400L80.eps}\hspace{0.2cm}\includegraphics[width=.49\textwidth]{plot-fluctdomain-k8-f-comp.eps}
\caption{Domain internal fluctuation as a function of the domain size. 
(a): data at fixed $f=0.2$ for different bending constants. 
(b): data at fixed $k=8$ for different fractions of reactive monomers.}
\label{fig:domfluct} 
\end{figure}

Following the procedure of Ref.~\cite{Moreno2016JPCL}, SCNP domains can be defined
as clusters of loops, where a loop is formed by all the monomers in the backbone contour between two cross-linked monomers. 
Two loops are merged into the same cluster if they share at least one monomer.
Fig.~\ref{fig:snap-domains} shows the structure of domains in two representative SCNPs with low and high asphericity. 
Clustering of loops is, in principle, a good criterion to define domains in SCNPs, since such clusters are expected to be tightly linked and, therefore, to be weakly deformable. This is confirmed for all the investigated systems in Fig.~\ref{fig:domfluct},
which shows the average relative fluctuation of the domain as a function of its size, defined as its number of monomers $n_{\rm d}$. 
The relative fluctuation is measured as $\delta = [(\langle R^2_{\rm g} \rangle - \langle R_{\rm g} \rangle ^2 )/\langle R^2_{\rm g} \rangle]^{1/2}$, with $R_{\rm g}$ the radius of gyration of the domain. 
As expected, the smallest domains (a few monomers) are close to the rigid limit ($\delta \rightarrow 0$). 
The relative fluctuation increases with the domain size, but the domains are still weakly deformable, saturating to a small value $0.05 < \delta < 0.12$. Not surprisingly, increasing $k$ and $f$ (i.e., chain stiffness and cross-linking density) leads to less deformable domains and lower values of $\delta$.

\begin{figure}[htb!]
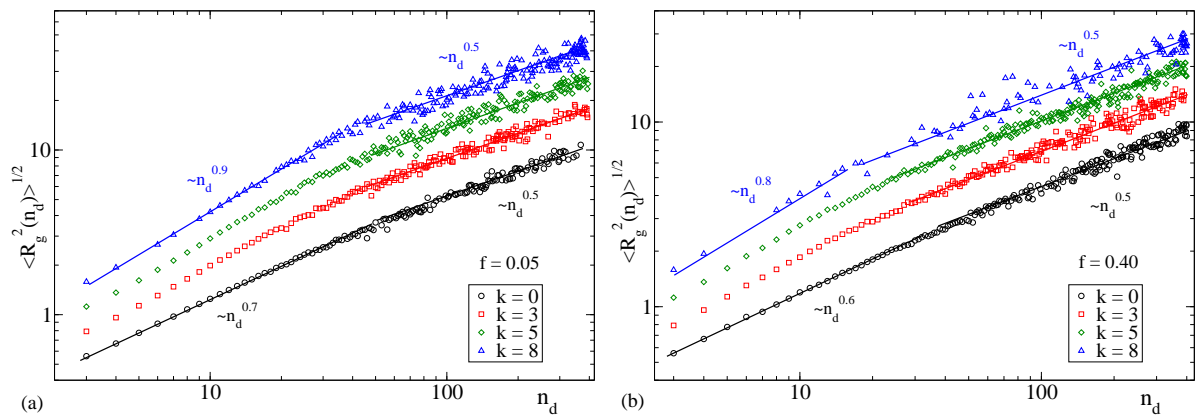

\centering
\includegraphics[width=.49\textwidth]{plot-rgdom-N400L20.eps}\hspace{0.2cm}\includegraphics[width=.49\textwidth]{plot-rgdom-N400L160.eps}
\caption{Radius of gyration of the domains as a function of their number of monomers. 
(a): data for $f=0.05$; (b): data for $f=0.40$. Different data sets correspond to different bending constants.
For the sake of clarity each set has been rescaled in the ordinate axis, from bottom to top,
by a factor 1, 2, 4 and 8. 
Lines are fits to power-law behaviour, $\langle R^2_{\rm g}(n_{\rm d}) \rangle  \sim n_{\rm d}^y$.
The exponents are indicated.}
\label{fig:rgdom} 
\end{figure}

Fig.~\ref{fig:rgdom} shows the domain radius of gyration vs. its number of monomers, for $f = 0.05$ and $f = 0.40$ at all the
investigated bending constants. Lines are fits to power-law behaviour 
$\langle R^2_{\rm g}(n_{\rm d}) \rangle  \sim n_{\rm d}^y$. Two different power-law regimes are found for small and big domains.
For small domains the exponent $y$ changes by increasing $k$ and $f$ from $y \gtrsim 0.59$ to $y \lesssim 1$, in a similar fashion to
self-avoiding chains/rings and rod-like objects, respectively.
Somewhat unexpectedly, in view of the trends found for other observables in the previous subsections,
the domains of size $n_{\rm d} > 50$ show in all cases an exponent $y \approx 0.5$, irrespective of the degree of cross-linking and chain stiffness. This exponent indicates that in all the investigated cases the SCNP domains are, in average,
weakly deformable but sparse objects. However, the analysis of the scaling properties 
of the whole SCNP (Figs.~\ref{fig:rs} and \ref{fig:wq}) suggests
that the chain stiffness and fraction of reactive monomers determines the particular spatial arrangement of the domains,
which can result in sparse or in globular arquitectures of the SCNPs.

\begin{figure}[htb!]
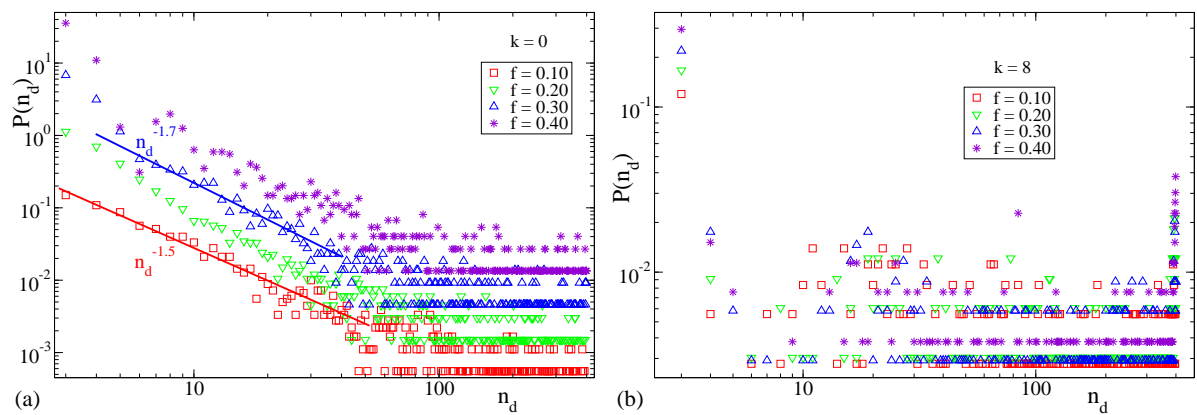

\centering
\includegraphics[width=.49\textwidth]{plot-distnclus-k0-f-comp-scal.eps}\hspace{0.2cm}\includegraphics[width=.49\textwidth]{plot-distnclus-k8-f-comp.eps}
\caption{Distributions of the domain size in SCNPs with $k=0$ (a) and $k=8$ (b). 
Different data sets in both panels correspond
to different fractions of reactive monomers. For the sake of clarity the sets in (a) have been rescaled in the ordinate 
axis by factors 1, 4, 16 and 64, from bottom to top. 
The lines indicate approximate power-law behaviour 
$P(n_{\rm d}) \sim n_{\rm d}^{-x}$ with $1.5 \leq x \leq 1.7$.}
\label{fig:dist-domain} 
\end{figure}

Fig.~\ref{fig:dist-domain} shows the distributions of domain sizes as a function of the fraction of reactive monomers for the fully-flexible ($k=0$) and the stiffest investigated SCNPs ($k=8$). Stiffness has a dramatic effect in the shape of the distributions. A monotonous decay is found for $k=0$, compatible with a power law $P(n_{\rm d}) \sim n_{\rm d}^{-x}$, with $1.5 \leq x \leq 1.7$. Similar monotonous decays
are found for $k=3$ and 5 (not shown), though with lower exponents ($x \sim 1.3$ and 1.0, respectively). 
A rather different behaviour is observed for $k=8$. After the initial decay the distribution becomes roughly flat. A peak at $n_{\rm d} \rightarrow N = 400$ arises for moderate and high cross-linking ratios ($f \geq 0.2$). This reflects a significant fraction of SCNP topologies consisting of a tightly linked single domain containing most of the monomers and
a few small dangling domains (as in Fig.~\ref{fig:snap-domains}a).
For example, 30\% of the SCNPs with $k=8$ and $f=0.3$ have a domain with 
300 of the total of $N=400$ monomers in the SCNP.

\begin{figure}[htb!]
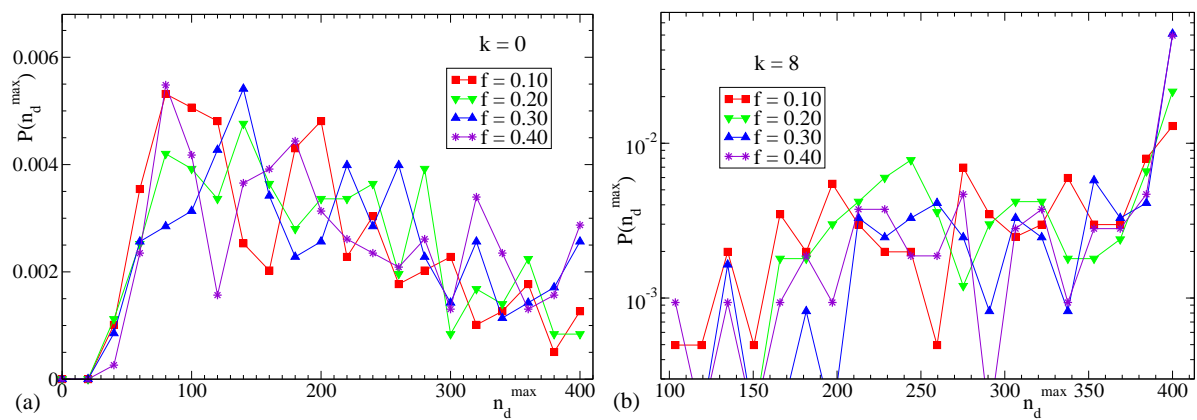

\centering
\includegraphics[width=.49\textwidth]{plot-clusmax-k0-f-comp.eps}\hspace{0.2cm}\includegraphics[width=.49\textwidth]{plot-clusmax-k8-f-comp.eps}
\caption{(a) Distribution of the size of the maximum domain in SCNPs with $k=0$ (a) and $k = 8$ (b).
The different sets correspond to different values of the fraction of reactive monomers.}
\label{fig:maxdom} 
\end{figure}

This effect can be emphasized by identifying the biggest domain in each SCNP, of size $n_{\rm d}^{\rm max}$ monomers, and representing the
corresponding distribution of $n_{\rm d}^{\rm max}$. This is shown in Fig.~\ref{fig:maxdom} as a function of $f$ 
for the fully-flexible ($k=0$) and the stiffest investigated SCNPs ($k=8$, note the logarithmic 
scale in the ordinate axis of panel (b)). 
As can be seen, the size
of the biggest domain in the fully-flexible SCNPs adopts all the values $n_{\rm d}^{\rm max} > 25$ and is broadly distributed, 
with a slow decay above $n_{\rm d}^{\rm max} \sim 200$. A very different behaviour is found for $k = 8$.
In every SCNP the biggest domain has at least 100 monomers, and the distribution $P(n_{\rm d}^{\rm max})$ is dominated by SCNPs
where the biggest domain contains most of the monomers (see the steep increase for $n_{\rm d}^{\rm max} > 350$).
Since a domain is defined as a cluster of loops, we conclude from the former results that stiff SCNPs are 
characterized by highly interconnected loops.

\section{Conclusions}
	
We have presented a detailed characterization of the conformational
properties of SCNPs as a function of the bending stiffness
of their linear polymer precursors. The broad range of investigated stiffness
has explored values from the limit of flexible chains to those characteristic
for stiff common polymers. Increasing stiffness hinders bonding
of groups separated by short contour distances and, in order to complete cross-linking, increases looping over longer distances.
This mechanism leads to the formation of more compact nanoparticles with a structure of highly interconnected loops.
The characterization of several structural observables as loop size, intramolecular distances and form factors has revealed 
a crossover in the scaling behaviour of the SCNPs by increasing stiffness.
The scaling exponents change from those characteristic for Gaussian chains or rings in $\theta$-solvents
in the fully flexible limit, to values resembling fractal or `crumpled' globular behaviour when the SCNPs become very stiff.
Still, the distribution of loop sizes retains the behaviour expected for Gaussian paths even for very stiff SCNPs,
since unlike in the flexible limit a high fraction of long loops is formed at the early and intermediate stage
of the cross-linking process, where the partially cross-linked precursor is still a self-avoiding object.

Increasing stiffness leads to bigger and less deformable SCNP domains.
Surprisingly, the scaling behaviour of the domains is in all cases
similar to that of Gaussian chains or rings, irrespective of their stiffness and degree of cross-linking.
It is the particular spatial arrangement of the domains which determines the global 
structure of the SCNP (sparse Gaussian-like object or crumpled globule).
Since intramolecular stiffness can be varied through the specific chemistry of the precursor or
by introducing bulky side groups in its backbone, the results presented in this article propose a new route to tune
the global structure of SCNPs.

\section{Acknowledgements}

We acknowledge financial support from the projects MAT2015-63704-P (MINECO-Spain and FEDER-UE) and IT-654-13 (Basque Government, Spain).
\\
\\
\\
\\
\\

\begin{thebibliography}{10}
\bibitem{notedefps}
It must be mentioned that this argument was originally invoked for a different
  definition of $P(s)$. In that case $s$ was defined as the countour distance
  between two monomers $i,j$ in contact, i.e. $r_{i,j} < r_0$, with $r_0 =
  \gamma r_{\rm cut}$ and $\gamma \lesssim 1$. We have computed $P(s)$ with
  this definition and reasonable values $0.9 < \gamma < 1$. No differences
  within statistics are found at middle and large $s$ with the function
  computed by using only the cross-linked monomers.
\end{thebibliography}

\providecommand{\newblock}{}

\end{document}